\documentclass[aps,prb,twocolumn,superscriptaddress,nobalancelastpage]{revtex4-1}
\usepackage[english]{babel}
\usepackage{graphicx}
\usepackage{amssymb,amsfonts,amsmath,mathtext,enumerate}
\usepackage{color}

\begin{document}
\title{Magnon spectrum in two- and three- dimensional skyrmion crystals}
\author{D.N. Aristov}
\affiliation{``PNPI'' NRC ``Kurchatov Institute'', Gatchina 188300, Russia}
\affiliation{Department of Physics, St.Petersburg State University, 7/9 Universitetskaya nab., 199034 St.~Petersburg, Russia}
\affiliation{Institute for Nanotechnology, Karlsruhe Institute of Technology, 76021
Karlsruhe, Germany }
\author{A.V. Tsypilnikov}
\affiliation{``PNPI'' NRC ``Kurchatov Institute'', Gatchina 188300, Russia}
\affiliation{Department of Physics, St.Petersburg State University, 7/9 Universitetskaya nab., 199034 St.~Petersburg, Russia}
\date{\today}

\begin{abstract}
We study the low-energy  magnon spectrum of the skyrmion crystal  (SkX) ground state, appearing in  
two-dimensional ferromagnet with Dzyaloshinskii-Moriya interaction and magnetic field. 
We approximate SkX hexagonal superlattice by a set of overlapping disks, and find the lattice period by minimizing the classical energy density. The determined spectrum of magnons on the disc of optimal radius is stable and only two lowest energy levels can be considered as localized. The subsequent hybridization of these levels in the SkX lattice leads to tight-binding spectrum. The localized character of the lowest magnon states is lost at small and at high fields, which is interpreted as melting of SkX. The classical energy of SkX  is slightly above the energy of a single conical spiral, and a consideration of quantum corrections can favor the skyrmion ground state. Extending our analysis to three-dimensional case, we argue that these quantum corrections become more important at finite temperatures, when the average spin value is decreased.  
\end{abstract}

\maketitle

\section*{Introduction}

Topological properties of the condensed matter systems is under active investigations over the last decade. 
Topological objects in magnetically ordered systems include 
various exotic spin structures, which are interesting both  theoretically and experimentally, see  \cite{SeidelBook2016} and references therein. One example is a so-called skyrmion crystal, which is a regular array of magnetic vortices.  The possibility of skyrmion textures  has been envisioned in the earlier works.\cite{Belavin1975, Bogdanov1994,   Ivanov1995, roslerBogdanov2006}

The existence of skyrmion lattices was  experimentally proven recently, in particular in the magnetic B20 compounds. \cite{PhysRevB.85.174416, Muhlbauer2009, Yu2010, Yu2010a, Grigoriev2007}  Nowadays materials with confirmed skyrmion lattice include, e.g., metals $\mathrm{Mn\,Si}$ and $\mathrm{Fe\,Ge}$, semiconductors $\mathrm{Fe_{1-x}\,Co_x\,Si}$ \cite{PhysRevB.81.041203, PhysRevB.89.064416}, and insulators $\mathrm{Cu_2\,O\, Se\, O_3}$.\cite{Nagaosa2013}

\begin{figure}[h]
\includegraphics[width=0.8\columnwidth]{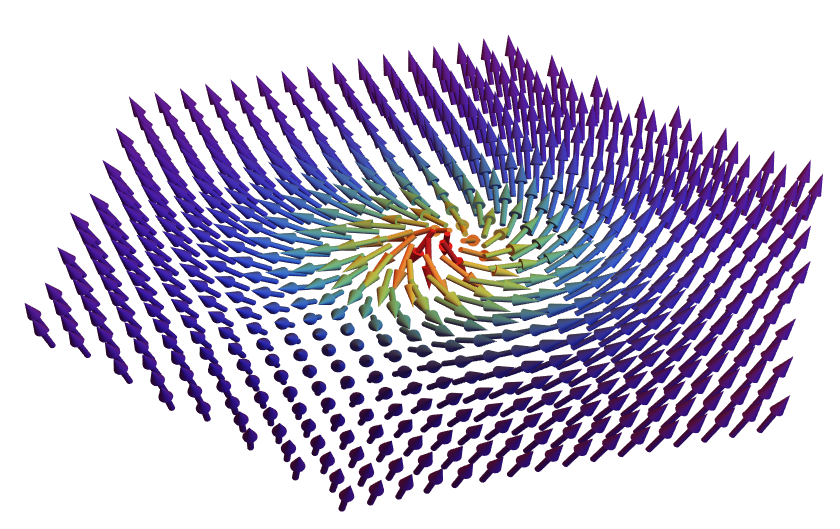}
\caption{A skyrmion configuration of localized magnetic moments  in a hexagonal cell.}
\label{pic:skyrmion}
\end{figure}

One of the specific property of one skyrmion (Fig.\ \ref{pic:skyrmion}) is its statical stability which is provided by topological protection and by the fact that this configuration of spins corresponds to a local minimum of the classical energy. \cite{Belavin1975} 

Topological protection of skyrmion configurations is interesting not only on the fundamental reasons, but also from a technical point of view. It is expected, that one can create a super-dense, long-term storage media \cite{Kiselev2011, Fert2013} and transistors \cite{Zhang2015} based on skyrmions. In addition, a skyrmion lattice strongly affects spin current. An electron moving through a skyrmion lattice changes its spin orientation multiply, in order to adjust it to a local magnetization vector. The skyrmion lattice causes an effective force, which changes an electron direction of motion that should manifest itself macroscopically as a kind of Hall effect. \cite{PhysRevB.94.024431} 

The static stability of skyrmion lattice was analyzed in \cite{Bogdanov1994}  in the classical limit, which limit is a usual theoretical approximation.  Similar approach was applied to various magnetic vortex structures \cite{metlov2002, metlov2010} and to arrays of magnetic dots. \cite{Galkin2006}  The semi-classical quantization method can be further used for the analysis of magnon spectrum, which was done for single skyrmion configuration in \cite{Schutte2014,Aristov2015, Aristov2016b}. 
Various aspects of  dynamics in  skyrmionic vortex structures were studied in \cite{Ivanov2002, Butenko2010, Dai2013, Ivanov1995, Ivanov1999}.

In this work we  investigate the low-energy spin-wave spectrum in the hexagonal skyrmion crystal.  Our analysis is done in three steps. We start by considering the two-dimensional (2D) magnet with exchange interaction, Dzyaloshinskii-Moriya interaction placed in the uniform magnetic field.  We study the system in classical limit and minimize its energy thus finding an optimal period of skyrmion structure.   A hexagonal unit cell of a skyrmion lattice is approximated by a disc  and  the semi-classical quantization is used to obtain magnon spectrum on it.  The energies and wave functions for magnons are found from the  Schr\"{o}dinger equation solved on the disc numerically.  This solution implies that we consider the hexagonal cells as isolated, by effectively applying infinite magnetic field on the borders, so that the magnon from one cell cannot pass to another. We relax this effective field in the further treatment by invoking the idea of small tunneling of low-energy magnons between cells. The amplitude of this tunneling is found from the form of the wave-function on the border, and we come to his the tight binding form of the spectrum \cite{book:abrikosov}. These findings are comparable to other studies of magnon spectrum on a skyrmion lattice. \cite{Mochizuki2012}

In addition to that we also discuss  effects connected to quantum nature of spin. First is the  zero-point motion of spins arising in the non-parallel spin configuration. This motion contributes to the ground state energy of the system, although this quantum correction is formally small by inverse value of spin, $s^{-1}$. We notice however, that the purely classical energy of skyrmion lattice is slightly higher than the energy of single conical spiral, and the quantum correction is needed to avoid metastability of skyrmionic state. 
At finite temperatures, $T$, the equilibrium value of spin decreases and the role of these quantum corrections in the stabilization of the skyrmion state should increase. 

Another effect is the quantum reduction of equilibrium spin value, which includes the zero-point contribution and the thermal correction. The zero-point contribution turns out to be small and non-uniform.
The consideration of finite $T$ is impossible in purely 2D case, as the spectrum is gapless and the fluctuations destroy the long-range magnetic order.  The inclusion of third spatial direction into our analysis is not difficult,  when we recall that the experimental evidence in B20 compounds   $\mathrm{Mn\,Si}$ and $\mathrm{Fe_{1-x} \,Co_x\, Si}$ shows the 3D skyrmionic structures  as stacks of two-dimensional lattices.  Within our model we show that the low-lying modes do not contribute much to the  thermal reduction of spin value, due to the small  volume of the Brillouin zone in the large-period skyrmion crystal.

The plan of the paper is as follows. We introduce the microscopic model and discuss its continuum limit in Sec.\ \ref{sec:lattice}. The basic ingredients of our analysis are introduced in Sec.\ \ref{sec:classics} and the determination of the superlattice period of skyrmion crystal is discussed in Sec.\ \ref{sec:radius}. The magnon spectrum on the disc with one skyrmion is discussed in Sec.\ \ref{sec:spectrum}. Particularly we analyze the dependence of the spectrum on radius of the disc  in Sec.\ \ref{sec:spectrumRadius},  quantum corrections to the ground state energy in Sec.\ \ref{sec:corrections} and  the quantum reduction of spin in Sec.\ \ref{sec:reduction}. The band structure of magnons in the SkX is evaluated in Sec.\ \ref{sec:bands} based on tight-binding model, the consistency of the model is checked here. The extension of our analysis to 3D case is done in Sec.\ \ref{sec:3d}. We present our conclusions in Sec.\ \ref{sec:conclusions}. The technical details of our derivation are given in three Appendices. 

\section{Skyrmion lattice model 
\label{sec:lattice}}

\subsection{2D chiral magnet and its classical description 
\label{sec:classics}}

We consider two-dimensional (2D) magnetic system without inversion center. In consideration we do not include anisotropy. \cite{Wilson2014} The model Hamiltonian on the square lattice is given by

\begin{equation}
\label{eq:ham}	
H = \sum\limits_{\langle ij\rangle} 
\left[ J_{ij} \, 
\widehat{\mathbf{S}}_{ i}\widehat{\mathbf{S}}_{ j}
 + \mathbf{D}_{ij} \cdot 
 \widehat {\mathbf{S}}_{i} \times \widehat {\mathbf{S}}_{ j} \right] - {\mathbf{B}} \sum\limits_{i }  \widehat{\mathbf{S}}_ {i}
\end{equation}
with $J_{ij} <0$ the ferromagnetic exchange, external magnetic field $\mathbf{B}$ is directed perpendicular to the 2D plane, and Dzyaloshinskii-Moriya (DM) interaction is characterized by the vector $\mathbf{D}_{ij} = D \left(\mathbf{r}_i - \mathbf{r}_j \right)$ is directed in the 2D plane.  \cite{Crepieux1998, Elhajal2002}

We   introduce the local magnetization, $\mathbf{m}(\mathbf{r}) = \langle \widehat{\mathbf{S}}_{\mathbf{r}} \rangle  $,  and  assume the semiclassical limit, $s\gg 1$. From this point onwards it is set $\hbar =1$. The exchange $J$ is regarded as the largest interaction, so that adjacent spins are almost parallel, it is convenient to subtract this large energy of uniform ferromagnet, $s^{2}\sum_{ij} J_{ij}$, from the subsequent consideration.  Making the gradient expansion of $\mathbf{m}(\mathbf{r})$, 
 as described  in Appendix \ref{app:discrete},  
we have the usual expression for the classical energy 
\begin{equation}
\label{eq:en_class}
E_{cl}= s^{2 }\int d\mathbf{r} \left( \frac{C}{2} \left( \nabla \mathbf{m} \right)^2 + D\mathbf{m}\left[ \nabla \times \mathbf{m} \right] - s^{-1}\mathbf{B}\, \mathbf{m} \right) \,, 
\end{equation}
with $C \sim |J| $ being spin stiffness constant.  When passing to continuum limit, i.e. $\sum_{i} \to a_{0}^{-2}\int d\mathbf{r}$, we introduce the dimension of length into the quantities $D \to D/a_{0}$ and $|\mathbf{B}|=B\to B/a_{0}^{2}$.  
In what follows we use dimensionless units. The energy is measured in units of $sC$   and the distance in units of $a_{0}C/D$.  The classical energy then has the large prefactor $s$ as compared to quantum Hamiltonian below without this factor. The remaining dimensionless parameter $b$ equals $BC/D^2s$. In  the discussion of the spectra below, we take $b=0.6$   because it produces the  results, typical for the general case.

One can find the static configuration of $\mathbf{m}(\mathbf{r})$ producing the energy minimum, and then also determine the equation of motion of fluctuations around it. The spectrum of these fluctuations corresponds to conventional spin waves.  However, if we want to retain a possibility to study effects of interaction between spin-waves, \cite{Aristov2016b} it is better to adopt another method described below.  Upon this the classical configuration and the spectrum of linear spin wave theory remain unchanged. 

Assume that the direction of the average magnetization $\mathbf{m}(\mathbf{r})$ changes with $\mathbf{r}$. 
We define the rotation matrix $\hat U\left( \mathbf{r}_{i} \right )$ at each site ${\mathbf{r}_i}$  so that $\mathbf{S}_{\mathbf{r}}  = \hat U\left( \mathbf{r} \right )  \tilde {\mathbf{S}}_{\mathbf{r}}$ and the average local spin   in the new basis, $\tilde {\mathbf{S}}_{\mathbf{r}}$,  is directed along the $\hat z$-axis. 
The position-dependent matrix is parametrized as $\hat U(\mathbf{r}) = e^{-\alpha \sigma_3}e^{-\beta \sigma_2}e^{-\gamma \sigma_3}$ with generators of $SO(3)$ group $\sigma_2$,$\sigma_3$ and Euler angles $\alpha$, $\beta$, $\gamma$. 
We expect that the average spins in new local bases are directed along $\hat z$, and we use the Maleyev-Dyson representation for spin operators, preserving the spin commutation relations  
\begin{equation} 
\begin{aligned} 
     \tilde{S}^{z}_{j} &=s-a^\dagger_{ j} a_{ j} \,, \quad   
       \tilde{S}^{+}_{j}=\sqrt{2s}a_{ j}  \\
     \tilde{S}^{-}_{j} &=\sqrt{2s}\left( a^{\dagger}_{ j} - \frac{1}{2s}a^\dagger_{ j}a^{\dagger}_{ j}a_{ j} \right)
  \end{aligned}  
  \label{eq:boz}
 \end{equation} 
 where $s$ is a value of spin, $\tilde S^{\pm} = \tilde S^{x} \pm i \tilde S^{y}$ and $[a_{ j},a^+_{ j}] = 1$. In the classical limit, $s\to \infty$, we  have $\tilde {\mathbf{S}}  =  s \hat z $ and \begin{equation}
\mathbf{m} = s \begin{pmatrix}
\sin \beta \cos \alpha  ,&  
\sin \beta \sin \alpha  , & 
\cos \beta 
\end{pmatrix} 
\label{eq:parametr}
\end{equation}

Before we proceed further, we outline our computation strategy.  
It is known that in a certain range of magnetic fields the Skyrmion crystal (SkX) may be formed, with the hexagonal superlattice characterized by some lattice spacing which is denoted $R$.
The average magnetization at the center of each hexagonal cell in perfect SkX is directed antiparallel to the field $\mathbf{B}$, and is parallel to it at the boundary of cells. 
To find a spectrum of magnetic excitations in such crystal we first determine an optimal $R$, denoted by $R_{0}$, from the minimization of the energy density. Next we find the magnon spectrum for elementary plaquette with the skyrmion at its center. Technically, these plaquettes in the form of discs \cite{Bogdanov1994, Bogdanov1994a, Bogdanov1989} are taken and then the whole plane is paved by these overlapping discs. The overlap between different plaquettes leads us to the tight-binding model on the hexagonal superlattice. Calculating the parameters of the latter model, the band structure for magnons in the whole system is obtained.

\subsection{Classical energy of skyrmion and optimal radius 
\label{sec:radius}}

\begin{figure}[t]
\includegraphics[width=0.8\columnwidth]{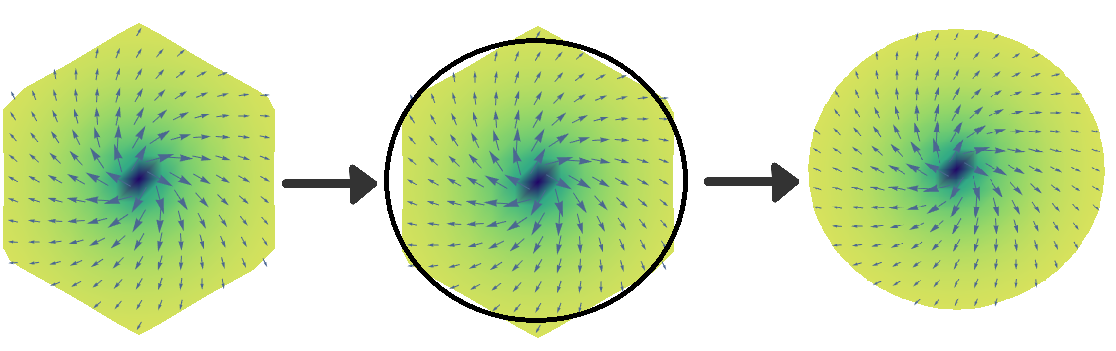}
\caption{ \label{pic:approxHexagone}
The hexagonal superlattice of skyrmion crystal is approximated by a set of overlapping discs. The radius of the disc, $R_0$, and hence the period of the superlattice is eventually determined by the energy minimization.}
\end{figure}

Our first step is to approximate the hexagonal unit cell by a disc of radius $R$. (Fig. \ref{pic:approxHexagone}). The optimal radius of the disc is found by minimization of  the average energy density for a single skyrmion $\rho_{c} = E_{sk}/ (\pi R^{2})$ on that disc. By doing this we minimize the energy of a whole skyrmion lattice \cite{Han2010}, which is approximatly equal to a sum of the energies of separate skyrmions (see below). 

For the analysis of a single skyrmion placed at the center of the disc of  radius $R$ we use polar coordinates $(x,y) = (r\cos \varphi, r \sin \varphi)$ and parametrize $\mathbf{m}$ by \eqref{eq:parametr}. 
Skyrmions  can be characterized by the  topological charge \cite{book:rajaraman}:
\begin{equation}
\begin{aligned}
Q & = \frac{1}{4\pi }
\int d\mathbf{r}\left( \mathbf{m}\left[ \frac{\partial \mathbf{m}}{\partial y} \times \frac{\partial \mathbf{m}}{\partial x} \right] \right)  \, , \\
& =   \frac1{4\pi}
\alpha \left( \varphi  \right) \big|_{\varphi  = 0}^{\varphi  = 2\pi }  \cdot 
 \cos \beta \left( r \right) \big|_{r = 0}^{ R}
\end{aligned}
\label{eq:topCharge}
\end{equation}
where the last line was obtained for centrosymmetric solutions $\alpha = \alpha(\phi)$ and $\beta = \beta (r)$.  
 
Spin configuration on the plane  can be viewed as mappings of one spherical surface onto another ($S^2 \rightarrow S'^2$) and can be classified into homotopy sectors.  \cite{book:rajaraman} 
Mappings within one sector can be continuously deformed onto another and there is infinity number of such homotopy sectors or classes,  characterized by integer $Q$. Non-trivial topological configuration of vector field corresponds to the boundary condition:
\begin{equation}
  {\alpha \left( \varphi  \right) = Q\varphi  + {\alpha _0}} \,, \quad 
  {\beta (0) = \pi } \,, \quad
  {\beta ({R}) = 0} \,.
\label{eq:edgeCond}
\end{equation}
Our solution corresponds to $Q=1$ and  $\alpha _{0} = \pm \pi/2$ as is shown below.   

The terms in Eq.\  \eqref{eq:en_class} can be represented in the form 
\begin{equation}
\begin{aligned}
  \left( \nabla \mathbf{m} \right)^2
  &= \left( \frac{d\beta }{dr} \right)^2 + \frac{\sin^2\beta }{r^2}
  \left( \frac{d\alpha }{d\varphi } \right)^2   \,,  \\
  \mathbf{m}\left[ \nabla  \times \mathbf{m} \right] 
  & = \sin \left( \alpha  - \varphi  \right)\left( \frac{d\beta }{dr} + \frac{1}{2r}\sin 2\beta   \frac{d\alpha }{d\varphi } \right)  \,,  \\
  \mathbf{B}\, \mathbf{m} &= B\cos \beta    \,.
\end{aligned}
\label{eq:ClassGlobalForm}
\end{equation} 
The exchange part of interaction, $  \left( \nabla \mathbf{m} \right)^2$, and  $\mathbf{B}\, \mathbf{m} $ are independent of the parameter $\alpha _0$. The DM part is related with topological charge and $\alpha _0$:
\[D \mathbf{m}\left[ \nabla  \times \mathbf{m} \right] =D \sin \left( \alpha _0 +(Q - 1 )\varphi  \right)\left( \frac{d\beta }{dr} + Q\frac{\sin 2\beta}{2r} \right)\]
The contribution of this term to classical energy for  $D > 0$  has a minimum at $Q=1$ and $ \alpha _0 = \pi/2  $.

\begin{figure}[t]
\includegraphics[width=0.9\columnwidth]{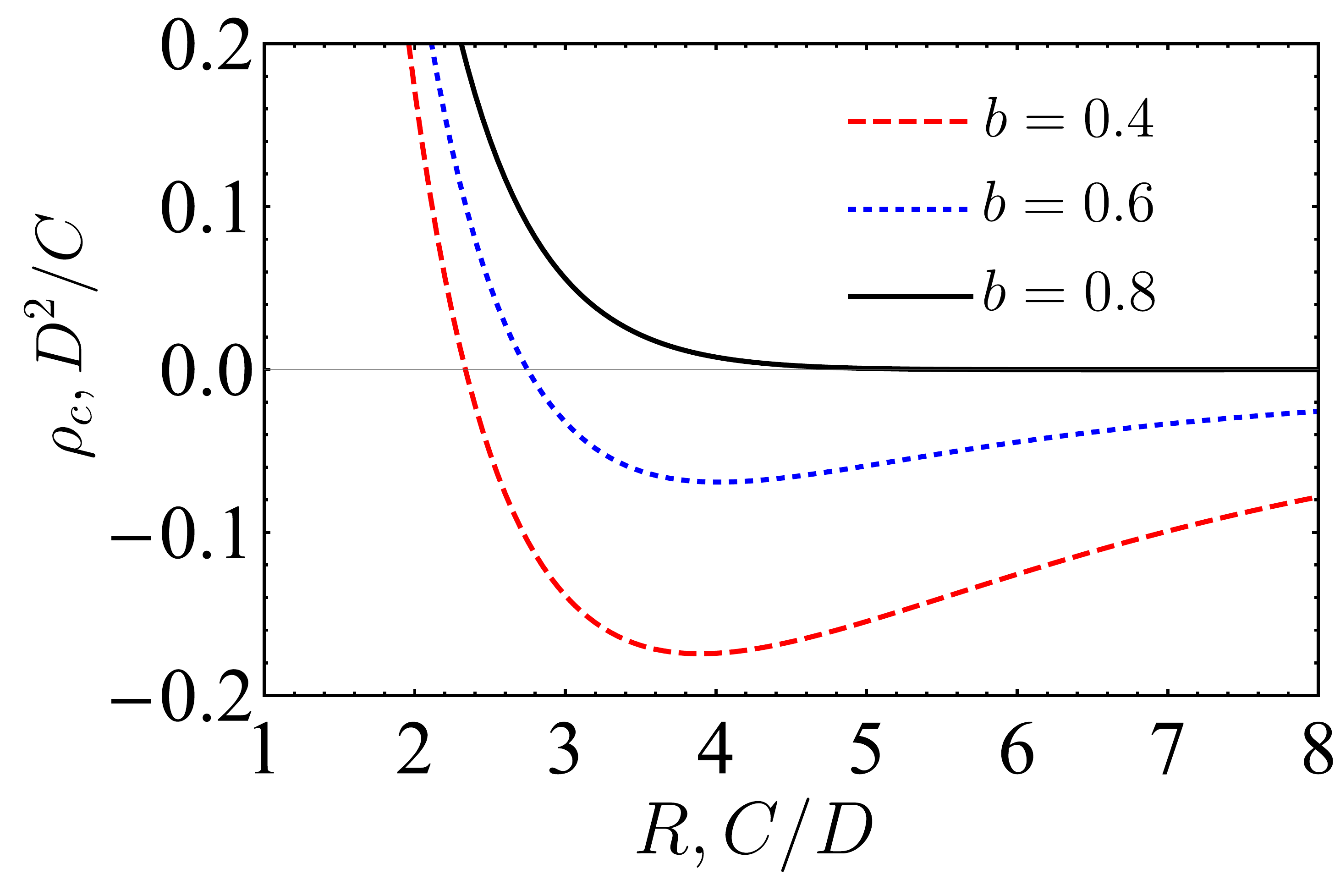}
\caption{\label{pic:plotDensity} 
Dependence of the energy density $\rho _c$  on the radius of the disc, $R$,  for different field strength, $b$.  The minimum of $\rho _c$ does not occur for   $b\agt 0.8$.}
\end{figure}

The classical energy \eqref{eq:en_class} is now evaluated with the use of  Eqs.\  (\ref{eq:edgeCond}),  \eqref{eq:ClassGlobalForm}, and dividing it by the area of disk $\pi R^{2}$ and by the factor $s$
\begin{equation}
\begin{aligned}
  \rho _c &=   \frac{2 }{R^2}\int _{0}^{R} 
  dr\left( \frac{\sin^2\beta }{2r} + \frac{r}{2}  \left( \frac{d\beta }{dr} \right)^2 \right.   \\
&\left. { + \left\{ r\frac{{d\beta }}{{dr}} + \frac{{\sin 2\beta }}{2} \right\} - br\left( {\cos \beta  - 1} \right)} \right)  
\end{aligned}
\label{eq:density}
\end{equation}
with the terms in curly brackets come from the DM part of interaction. For   convenience we subtracted  here the energy of uniform ferromagnet  so that  $\rho_{c} =0$ for $\beta \equiv 0$. 
The resulting Euler-Lagrange equation
\begin{equation}
\frac{d^2\beta }{dr^2} + \frac{1}{r}\frac{d\beta }{dr} - \frac{\sin \beta \cos \beta }{r^2} + \frac{2\sin ^2\beta }{r} - b\sin \beta  = 0 \,,
\label{eq:eulerLagr}
\end{equation}
is supplemented by the boundary conditions from \eqref{eq:edgeCond}.
The latter equation  is not of hypergeometric type thus its solution cannot be expressed in a closed form through hypergeometric functions. However it is readily solved numerically for any particular $R$ by shooting method. Proceeding this way,  we get the profile $\beta(r)$ and the dependence of $\rho_c$  on the  disc radius, $R_{0}$. The results for the energy density are presented in the Fig. \ref{pic:plotDensity}. 

It can be seen in the Fig.\ \ref{pic:plotDensity} that in the range $b \in (0, 0.8)$ there is a minimum in the energy density at some optimal radius, $R_{0}$. For magnetic field with larger values, $b \agt 0.8$,  the energy of skyrmion for any $R$  is higher than the energy of ferromagnetic ground state,  $\beta \equiv 0$.  We show below that the  value of optimal $R_{0}$ determined from the classical energy can be further refined by consideration of quantum corrections to the ground state.

\begin{figure}[t]
\includegraphics[width=0.95\columnwidth]{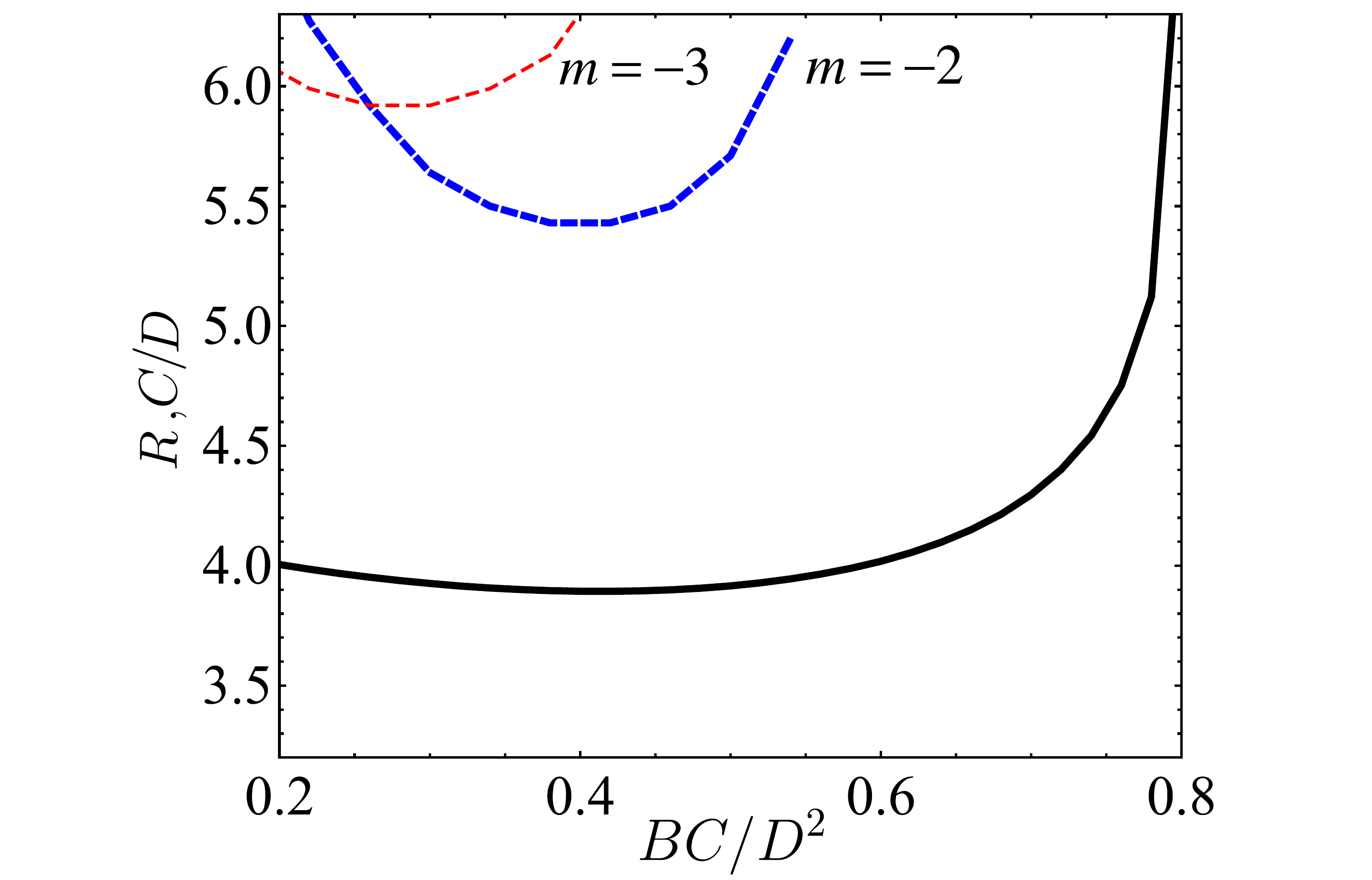}
\caption{ The solid line shows the value of optimal radius of the disc, $R_{0}$, providing the minimum of the classical energy density. The dashed curves above $R_0$ correspond to instabilities of skyrmion solution, discussed below. It means, e.g., that the energy of mode $\varepsilon_{-2,0}$ of skyrmion with $b \approx 0.5$ on the disk with $R\approx 5.6$ become negative, as shown in Fig\ \ref{pic:instability}.}
\label{pic:R0_field}
\end{figure}

We also plot the dependence of the determined $R_{0}$ on the magnetic field $b$ in the Fig.\  \ref{pic:R0_field}.    We obtain $R_{0} \sim 4$ almost in the whole range of $b$.  This indicates that $R_{0}$ is mostly determined by the ratio $C/D$ in accordance with below estimates for the period of single spiral.   Notice  that $R_{0}$ mildly diverges at $b_{c}\simeq 0.8$, in accordance with \cite{Schutte2014, Lin2014},  presumably by logarithmic law. 

\section{Magnon spectrum on the disc
\label{sec:spectrum}}


The above minimization of the classical energy happens in order $s^{2}$. The obtained dependence $\beta(r)$ is used in the next terms of the expansion of the Hamiltonian in powers of $s^{-1}$.  Each boson operator in this expansion is accompanied by the factor $s^{-1/2}$. The linear -in-bosons terms, $\sim s^{3/2}$, are absent as it should be for the extremum configuration in the semiclassical expansion. 
The  linear-in- $s$ terms are quadratic in bosons, and they constitute the linear spin-wave theory (LSWT) \cite{Schutte2014}:

\begin{equation}
H_{2}= \int d\mathbf{r} 
\left(   a_{\mathbf{r}}^\dag \hat F(\hat  L_{z} ) {a_{\mathbf{r}}} 
+ \tfrac{1}{2} \left( 
 G \, a_{\mathbf{r}}^\dag a_{\mathbf{r}}^\dag  + H.c. \right) 
\right)
\label{eq:HamQuantBose}
\end{equation}
with 
\begin{equation} 
\begin{aligned} 
  F  (\hat  L_{z} )&=    - \nabla^{2}  + \frac{1 + 3\cos 2\beta }{4r^2} - \frac{2\cos \beta }{r^2}\hat{L}_{z} - \frac{1}{2}\left( \frac{d\beta }{dr} \right)^2  \\
& +  \left\{  - \frac{3\sin 2\beta }{2r} + \frac{2\sin \beta }{r}\hat{L}_{z}  - \frac{d\beta }{dr} \right\} + b\cos \beta   \\ 
G &= \frac{1}{2}\left(  - \frac{\sin ^2\beta }{r^2} + \left( \frac{d\beta }{dr} \right)^2 \right) 
+ \left\{ \frac{d\beta }{dr} - \frac{\sin 2\beta }{2r} \right\}
  \end{aligned}  
 \end{equation} 
where  $\hat  L_{z} =  - i\frac{\partial }{\partial \varphi }$
and
$\nabla^{2}   = \frac{1}{r}\frac{\partial }{\partial r}\left( r\frac{\partial }{\partial r} \right) - \frac{\hat{L}_{z} ^2}{r^2}$.  Here again the curly brackets contain the DM contribution and the integration in \eqref{eq:HamQuantBose} is done within the disc $r<R$.

The local magnon operators can be expanded in  
the basis functions  $ e^{im\varphi } f_{n}^{(m)}( r )$ forming an orthonormal set on the disk
\begin{equation}
\begin{aligned}
  & a_{\mathbf{r}} = \frac{1}{\sqrt {2\pi } }
\sum_{m}  \sum\limits_{n\ge 0} e^{im\varphi } f_{n}^{(m)}( r )\, a_{m,n} , \\ 
  & \left[ a_{m,n}  ,a_{m',n'}^\dag  \right]  = \delta _{mm'}\delta _{nn'} 
\end{aligned}
\label{eq:expansionBose}
\end{equation}
 At the center of the disc  we have $F(\hat L_{z} =m) = 
- r^{-1} \partial_{r}  r  \partial_{r}  + (m+1)^{2} /r^{2}  + {\cal O}(1) $ and $G =  {\cal O}(1)$.  It is  therefore convenient to choose the Bessel functions with shifted index, $f_{n}^{(m)}( r ) \propto J_{m+1} (\kappa_{n} r)$, as discussed in Appendix  \ref{app:num}.   Notice that the normalization of the wave-functions introduces the square of the inverse length scale into the energies. It means that magnon energies are obtained in units of $D^{2}/C$.
 
Operators $a_{m,n}$  can be further represented in terms of true magnon operators $c_{m,k}$ via Bogoliubov transformation 
\begin{equation}
\begin{aligned}
  a_{m,n} &= \sum_{k\geq0} (  u^{(m)}_{nk} c_{m,k} + v^{(m)}_{nk} c^{\dagger}_{-m,k}   )
\end{aligned}
\label{eq:uvBogo}
\end{equation}
where we require 
\begin{equation}  
\begin{aligned}
\sum_k \big( u^{(m)}_{nk} u^{(m)}_{n'k} - v^{(m)}_{nk} v^{(m)}_{n'k}  \big )&= \delta _{nn'}   \\
\sum_k \big(u^{(m)}_{nk} v^{(-m)}_{n'k} - v^{(m)}_{nk}u^{(-m)}_{n'k} \big ) &= 0   
\end{aligned}
\label{eq:uvBogolubov}
\end{equation}
so that $ [c_{m,k} ,c_{m',k'} ^\dag  ] = \delta _{mm'}\delta _{kk'}$.

 In terms of these coefficients 
 we come to the eigenvalue problem   for each $m$
\begin{equation}
\sum_{k\geq0}^{\infty}
 \begin{pmatrix}
F_{nk} (m)&G_{nk}\\
 - G_{kn} & - F_{nk} (-m)
 \end{pmatrix}
\begin{pmatrix}  u_k^{(m)}\\    v_k^{(-m)}  \end{pmatrix}  
 = \varepsilon^0_{m,n}  \begin{pmatrix}  u_n^{(m)}\\    v_n^{(-m)}  \end{pmatrix} 
\label{eq:numMatrixEq}
\end{equation}
where  the expressions for $F_{nk} (m)$ and $G_{nk}$ are given in Appendix  \ref{app:num}.   

Alternatively we may define 
\begin{equation}  
\begin{aligned}
&a _{\mathbf{r}} =  
\frac{1}{\sqrt{2 \pi}}   \sum_{m}  \sum\limits_{k\ge 0} e^{im\varphi } ( \phi_{k}^{(m)}( r )\, c_{m,k}  + 
{\tilde \phi}_{k}^{(m)}( r )\, c^{\dagger}_{-m,k} ), \\ 
 &\phi_{k}^{(m)}( r )  = \sum_{n} f_{n}^{(m)}( r ) u^{(m)}_{nk}  , \quad 
{\tilde \phi}_{k}^{(m)}( r )   = \sum_{n} f_{n}^{(m)}( r ) v^{(m)}_{nk} , \\ 
\end{aligned} 
\label{uvfunction}
\end{equation}
which results in the equation 
\begin{equation}
\begin{pmatrix} F( m )-  \varepsilon^0 _{m,n} ,&G \\ G,& F(  - m ) +\varepsilon^0 _{m,n} \end{pmatrix}
\begin{pmatrix} \phi_{n}^{(m)}\\ {\tilde \phi}_{n}^{(-m)} \end{pmatrix} = 0 \,.
\label{eq:ShrodEq}
\end{equation}


\subsection{Spectrum dependence  on the disc radius
\label{sec:spectrumRadius}}

\begin{figure}[t]
\includegraphics[width=0.95\columnwidth]{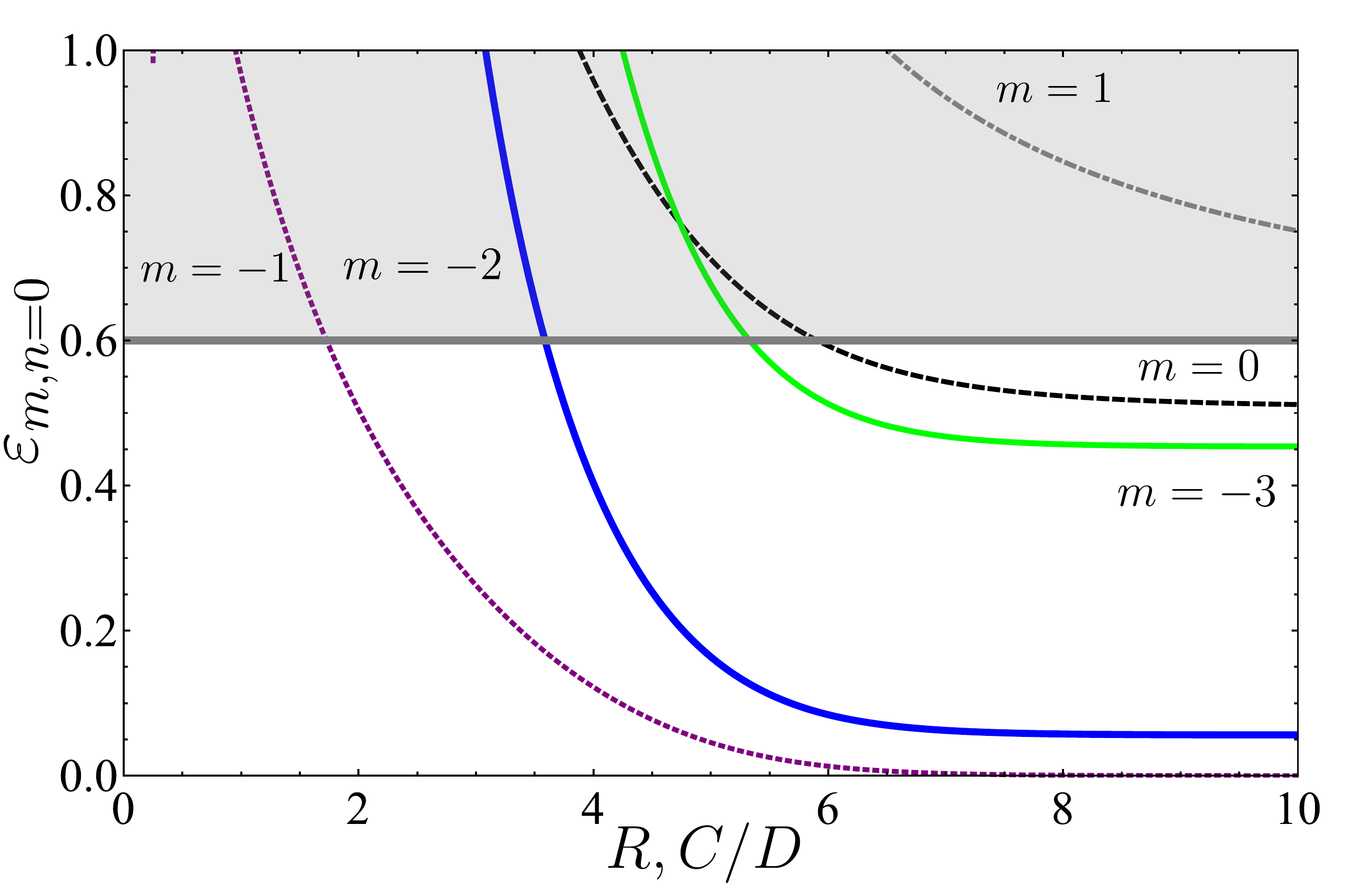}
\caption{Dependence of the energy of lowest modes on the disc's radius $R$ for $b=0.6$. There is only one  mode with $m=-1$, corresponding to infinitesimal translations, which has zero energy at large $R$. At optimal disc radius, $R_{0}\simeq 4$, the energy of only two modes are below the expected magnon continuum in the bulk, depicted as shaded area at $E>b = 0.6$.}
\label{pic:plotModeEnerg}
\end{figure}

\begin{figure}[t]
\includegraphics[width=0.9\columnwidth]{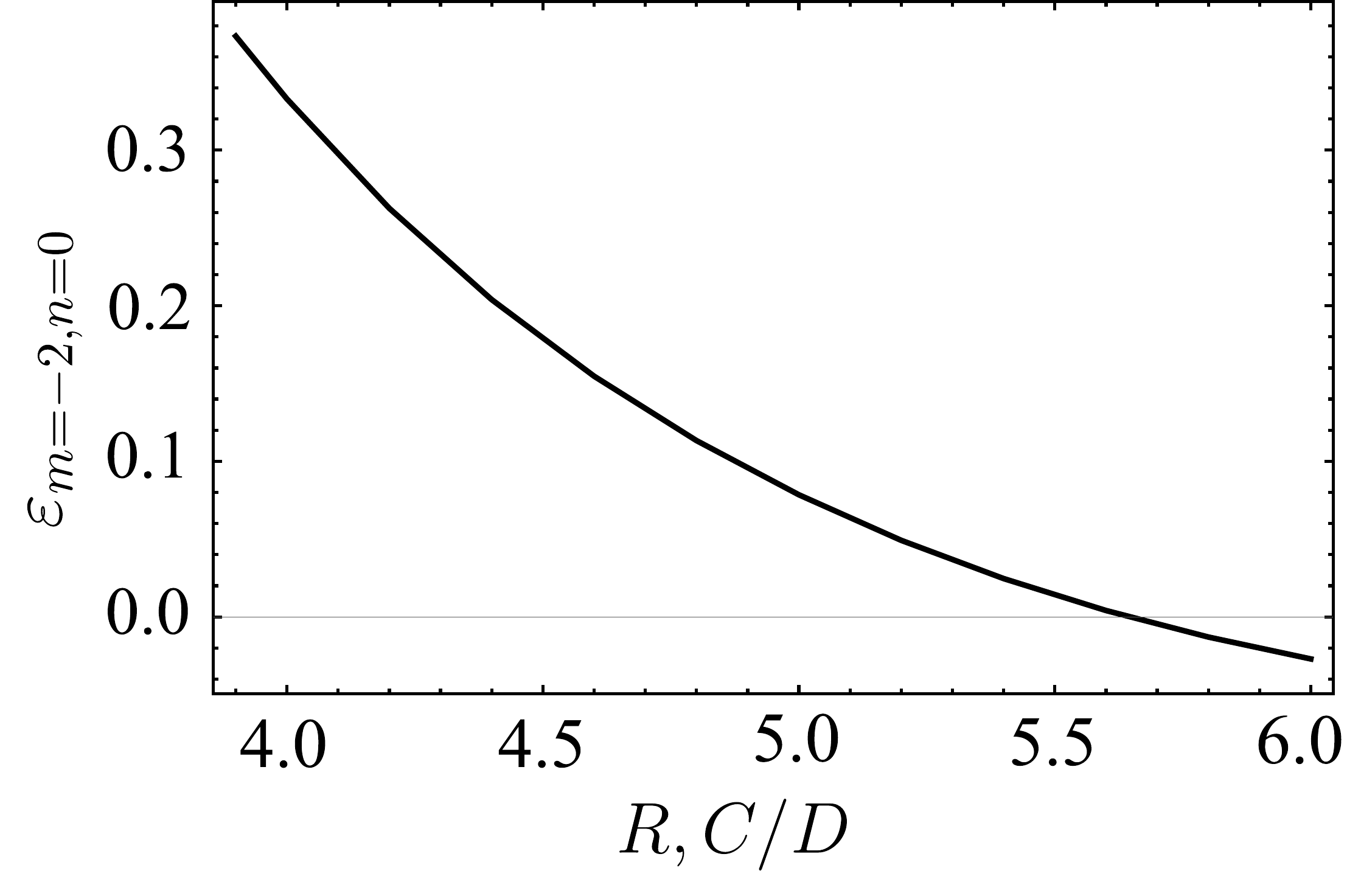}
\caption{ The dependence of the energy $\epsilon_{m=-2,0}$ of the lowest  mode with $m=-2$ on the radius of the disc at $b=0.5$. For optimal radius, $R_{0}=3.9$, the energy is $\epsilon _{-2,0} =0.37$, whereas for large values $R\agt 5.6$ the energy becomes negative, indicating instability of the system\cite{Schutte2014,Ezawa2011}. }
\label{pic:instability}
\end{figure}

Using the method outlined above we computed the magnon spectrum on the disc for various $m$.  
To show a wider picture, we first relax the condition $R=R_{0}$ and depict the dependence of the energy levels on $R$ in the Fig. \ref{pic:plotModeEnerg}.
For better visibility we show for each $m$ the lowest level for radial quantization. 

It was shown in \cite{Aristov2015} that in the absence of DM interaction and of magnetic field the magnon spectrum for Belavin-Polyakov skyrmion possesses three zero modes, corresponding to conformal symmetries of the action. In the present notation these modes have $m=-1,0,1$, for translation, dilatation and special conformal (rotation at infinity) symmetry, respectively.  The modes with $m=0$ and $m=1$ acquire the finite energy due to existence of DM interaction and magnetic field.   The energy of the mode with  $m=0$ is non-zero because the DM interaction defines the characteristic scale, $C/D$. The mode with $m=1$ has positive energy because the direction of magnetic field breaks the rotational symmetry in spin space.

The mode with $m=-1$ has  non-zero energy because the translational invariance is lost on the disc. Its energy is comparable to other modes at optimal radius, $R_{0} \simeq 4$, and decays exponentially with $R$ cf.  \cite{Schutte2014}. All energies tend to constant values at large $R$. 
The formal expression existing  for zero mode $\psi_{m=-1,n=0}$  at $R\to \infty$ (see Eq.\ (51) in\cite{Schutte2014})  is not inapplicable to our case   of finite $R$   because $\beta'(R) \neq 0$.

The obtained spectrum is always stable, i.e. all energies are positive, for optimal disc radius, $R_{0}$. This was checked for various $b$ and should be compared with the instability, discussed  in \cite{Schutte2014}. Namely,  negative energies were obtained for $m=-2$,  $b\alt 0.56$ and for large $R$.  The latter instability for a disc with non-optimal $R\neq R_{0}$ indicates that the energetically favorable spin configuration is not a single skyrmion ground state. \cite{Schutte2014,Ezawa2011} We show the calculated lowest energy for $m=-2$, $b=0.5$ and various $R$ in the Fig.\ \ref{pic:instability}. In this case $\epsilon _{-2,0} >0$ for optimal value $R=R_{0}=3.9$ and becomes negative at larger $R\agt 0.56$.

\subsection{Quantum correction to the ground state energy
\label{sec:corrections}}

The LSWT Hamiltonian \eqref{eq:HamQuantBose} is normal-ordered in 
$a_{\mathbf{r}}^\dag$, $ a_{\mathbf{r}}$. The diagonalization of it in matrix form is done by first 
representing 
\[  H_{2} =   \frac 12 \int d\mathbf{r}  \, 
  ( a_{\mathbf{r}}^\dag ,    a_{\mathbf{r}} ) 
\begin{pmatrix}
\hat F  ,&   G  \\ 
 G  , & \hat F 
\end{pmatrix}
 \begin{pmatrix}
  a_{\mathbf{r}} \\   a_{\mathbf{r}} ^\dag 
 \end{pmatrix}
 \] 
then expressing it in the basis \eqref{eq:numMatrixEq} and subsequently finding the unitary Bogoliubov transformation to come to the form $H_{2} = \sum _{mn}  \varepsilon^0_{m,n} c_{mn}^{\dag}  c_{mn}$, which is diagonal in terms of true magnon  operators, $ c_{mn}^{\dag} $, $ c_{mn} $.  It is easily verified that these manipulations involve a loss of normal ordering in terms of $a_{\mathbf{r}}$, and returning to normal ordering in terms of  $ c_{mn}$. Two appearing commutators are not equal and their difference is the quantum correction, $E_{q}$, to the energy of the ground state. 
After some calculation we find
\begin{equation}
E_{q} = \frac{1}{2}\left( \sum_{m,n} \varepsilon^0_{m,n} - \sum_m \mbox{Tr }  F_m  \right) \,.
\label{eq:Q_correction}
\end{equation}
This quantity is calculated on the disc of the radius $R$ using the formulas in Appendix \ref{app:num}. The results are shown in  Fig.\  \ref{pic:qCorEn} for the energy and its average density, $\rho_{q} = E_{q}/\pi R^{2}$. 
The dependence of density of quantum correction, $\rho_{q}$ on $R$ is rather pronounced and    $\rho_{q}\to 0$  at large $R$. What is interesting is the existence of the minimum at   $R \sim 1$,  where the density of classical (ground state) energy is high, see Fig.\ \ref{pic:plotDensity}.

\begin{figure}[t]
\includegraphics[width=0.95\columnwidth]{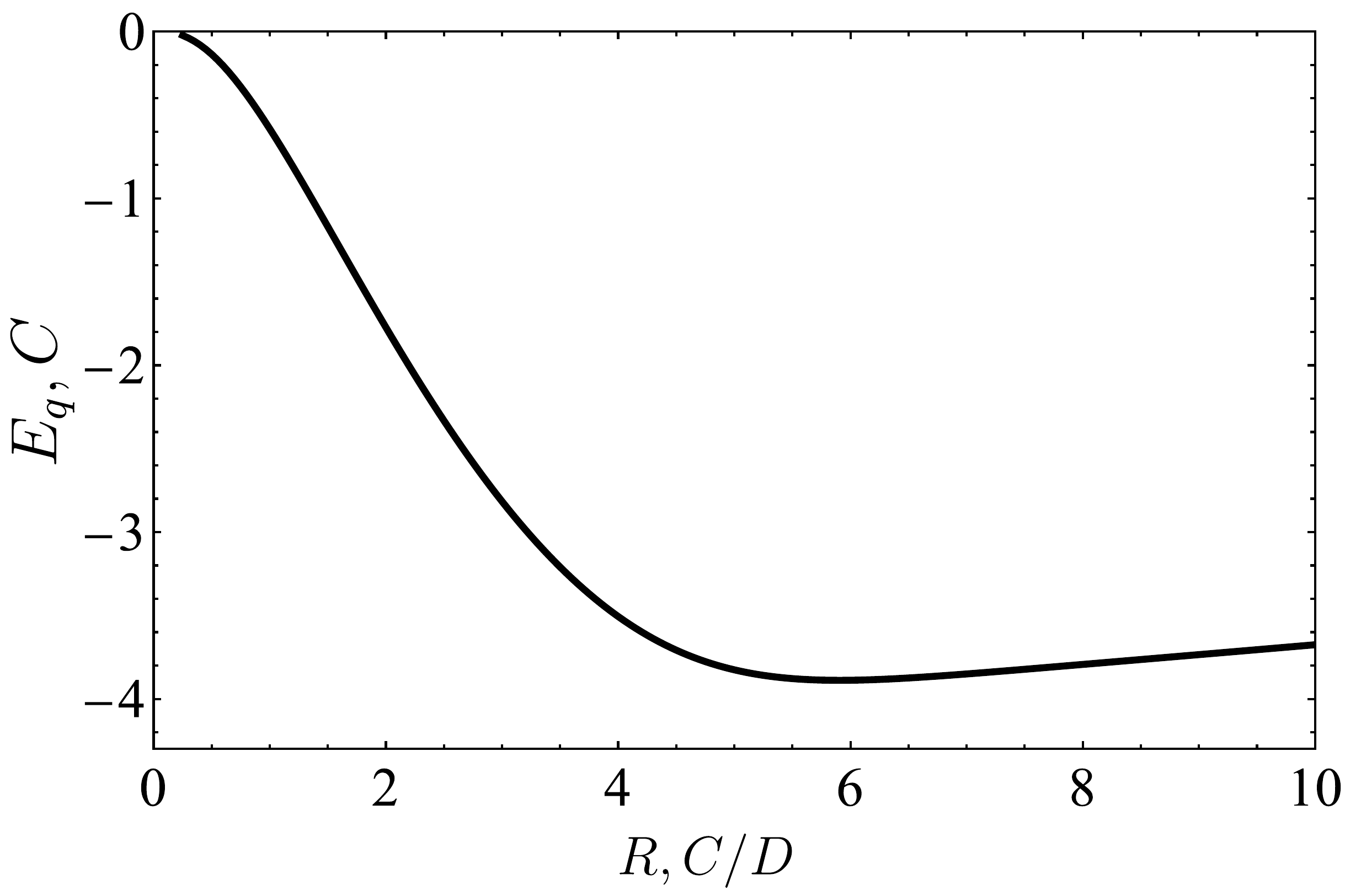}
\includegraphics[width=0.95\columnwidth]{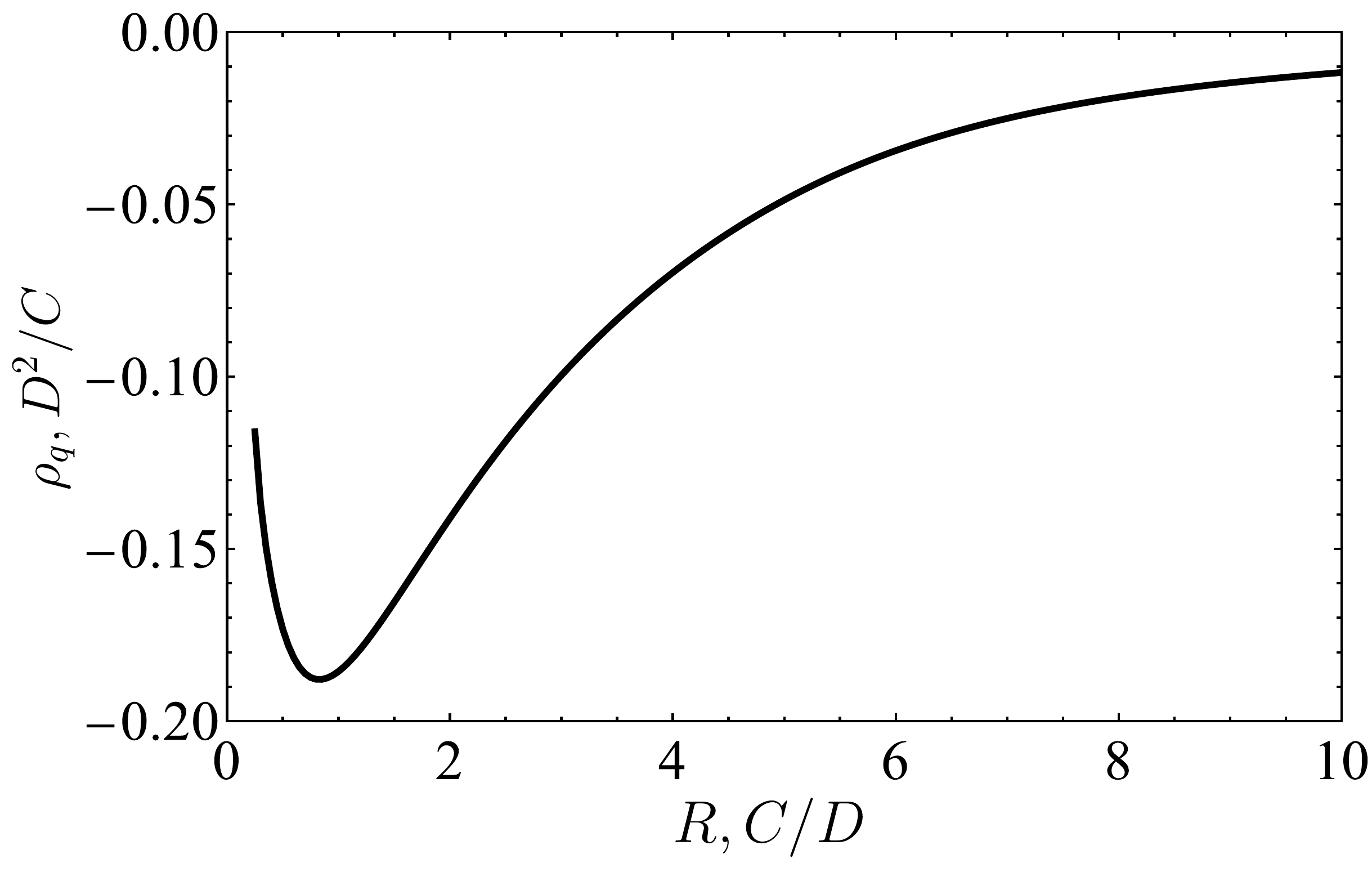}
\caption{The dependence of energy of quantum correction and its density on a radius of a disc. In contrast to Fig.\  \ref{pic:plotDensity} the minimum of energy density  is not at $R_0 \approx 4$.}
\label{pic:qCorEn}
\end{figure}

\begin{figure}[t]
\includegraphics[width=0.95\columnwidth]{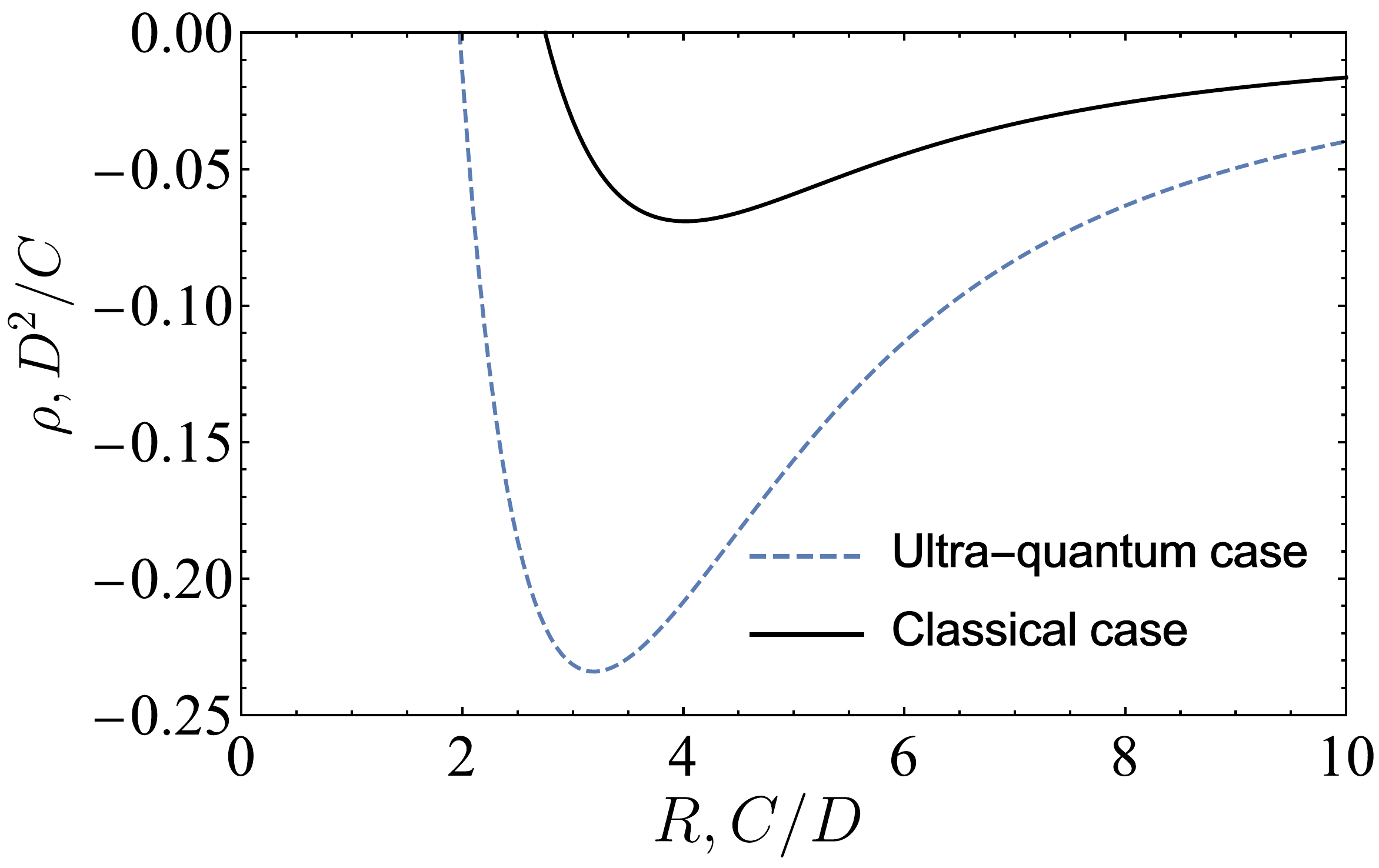}
\caption{The energy density with quantum corrections in ultra-quantum case ($s=1/2$) shown in comparison with the classical energy density ($s\to \infty$) . The curves are normalized by the factor $s$.  We see a change in the position of  the minimum towards a  new optimal radius $R_{opt}$, see text.}
\label{pic:ultraQuantCase}
\end{figure}

It means that the above optimal radius, $R_{0}$, calculated purely in classical terms, does not correspond to the minimum of energy density at finite $s$. The consideration of quantum correction shifts the optimal $R_{0}$ to smaller values.  This correction can be neglected in case $s \gg 1$, where our theory should work well.    
It is instructive to evaluate the total ground state energy, $E_{cl} + E_{q}$, for the  ultra-quantum case $s=1/2$. 
As shown in the Fig.\ \ref{pic:ultraQuantCase} the optimal radius in this latter case becomes $R^{uq}_{opt} \simeq 3.5 < R_{0} \simeq 4$. Summarizing, we see that the consideration of the quantum correction to the ground state energy  lowers the optimal values for the skyrmion crystal lattice spacing, but this shift in $R_{0}$ is not too big even for small $s$. Therefore we can safely use the previously determined spectrum, Fig.\ \ref{pic:plotModeEnerg}, which exhibits no big variation around $R=R_{0}\simeq 4$. 

\begin{figure}[t]
\includegraphics[width=0.95\columnwidth]{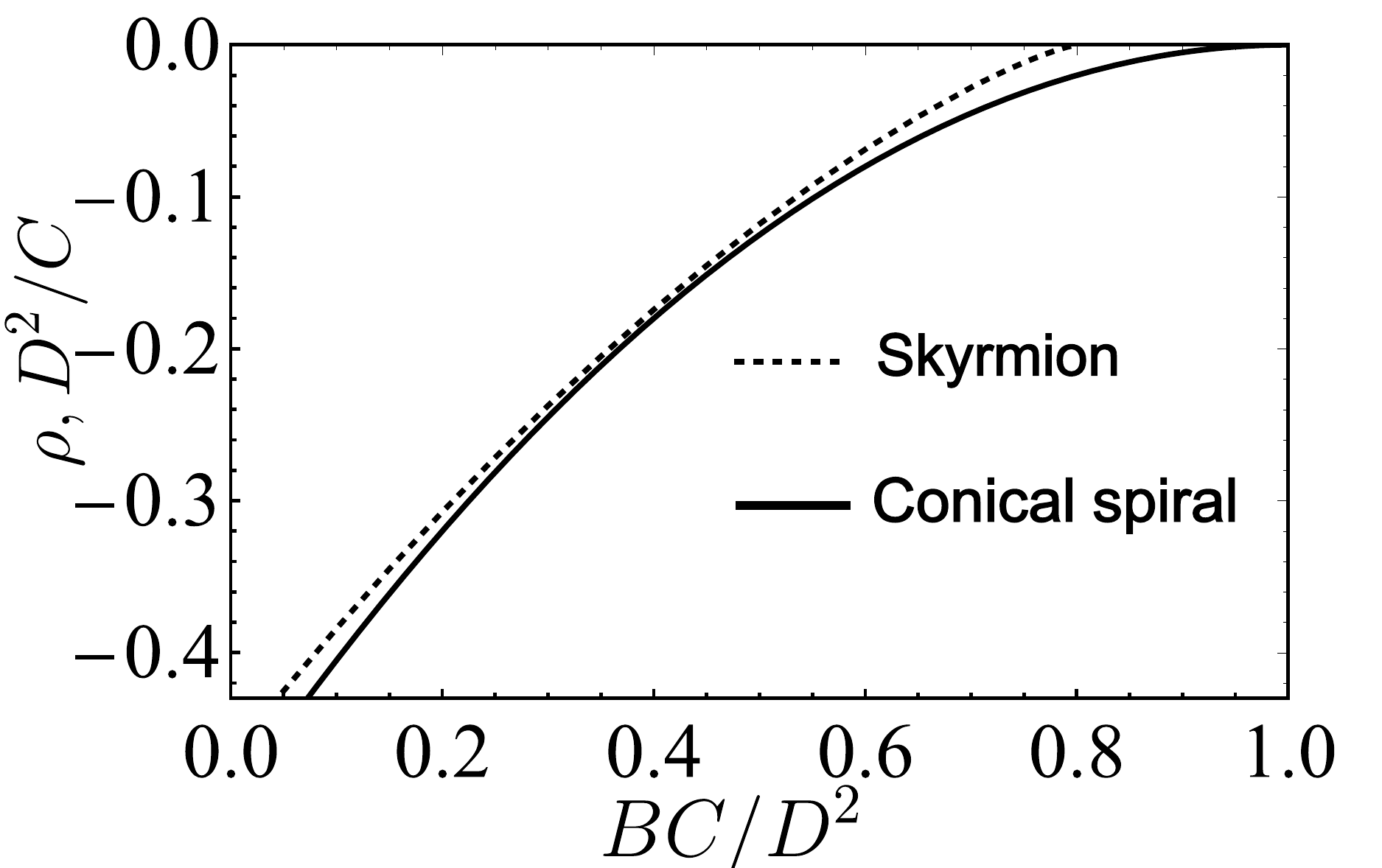}
\caption{The classical energy denisty of the skyrmion state (dotted line, Eq.\ \eqref{eq:density}) shown in comparison with the energy of conical spiral (solid line, Eq.\ \eqref{spiral_energy}). The curves are normalized by the factor $s$.  }
\label{fig:Sk_v_spiral}
\end{figure}

It can be argued \cite{Maleyev2006} that the single conical spiral is a good candidate for the true ground state of the  system, described by Eq.  \eqref{eq:en_class}. As we show in Appendix  \ref{app:spiral}, the energy density of the conical spiral without the ferromagnetic  contribution, $-B$, is given by the expression 
 \begin{equation}
\begin{aligned}
\rho_{sp} & = - \frac{D^{2}}{2C} (b-1)^{2} \,, \quad  0<b<1   \\
 &= 0   \,, \quad  b \ge1  \\
\end{aligned}
\label{spiral_energy}
\end{equation}
In the Fig.\ \ref{fig:Sk_v_spiral} we compare the energy density 
described by Eq. \eqref{spiral_energy} with the density determined for the skyrmion configuration on the disc, Eq.\ \eqref{eq:density}. We see that in the whole range of the fields, $b\in (0,1)$ the energy of the conical spiral is slightly lower than that of the skyrmion configuration. This means that the skyrmion configuration should be considered as metastable.  Although this fact does not prevent us from the determination of the magnon spectrum below, it may cast doubts in such procedure. We   note here that the discussed quantum corrections to the ground state   lowers the energy of the skyrmion configuration and can ultimately make it favorable. The validation of the latter statement requires however a calculation of the quantum correction to the spiral state, which is beyond the scope of the present study. 

\subsection{Quantum spin reduction
\label{sec:reduction}}

In addition to quantum correction to the ground state energy we can discuss the contribution of the zero point motion to the equilibrium spin value, $\langle \delta S^{z} _{\mathbf{r} } \rangle = s - \langle \tilde S^{z}  _{\mathbf{r} }\rangle = \langle a^{\dagger} _{\mathbf{r} } a _{\mathbf{r} }\rangle $.  It is well known that this correction in case of uniform 2D Heisenberg antiferromagnet on the square lattice is $\langle \delta S^{z} _{\mathbf{r} } \rangle \simeq 0.20$, because the staggered magnetization is not a conserved quantity.  In our case this quantum spin reduction appears also because the sum of equilibrium vector spin values, $\sum_{\mathbf{r}} \mathbf{m} (\mathbf{r}) \mathbf{S}_{\mathbf{r}}$, does not correspond to the conserved operator. 

\begin{figure}[t]
\includegraphics[width=0.95\columnwidth]{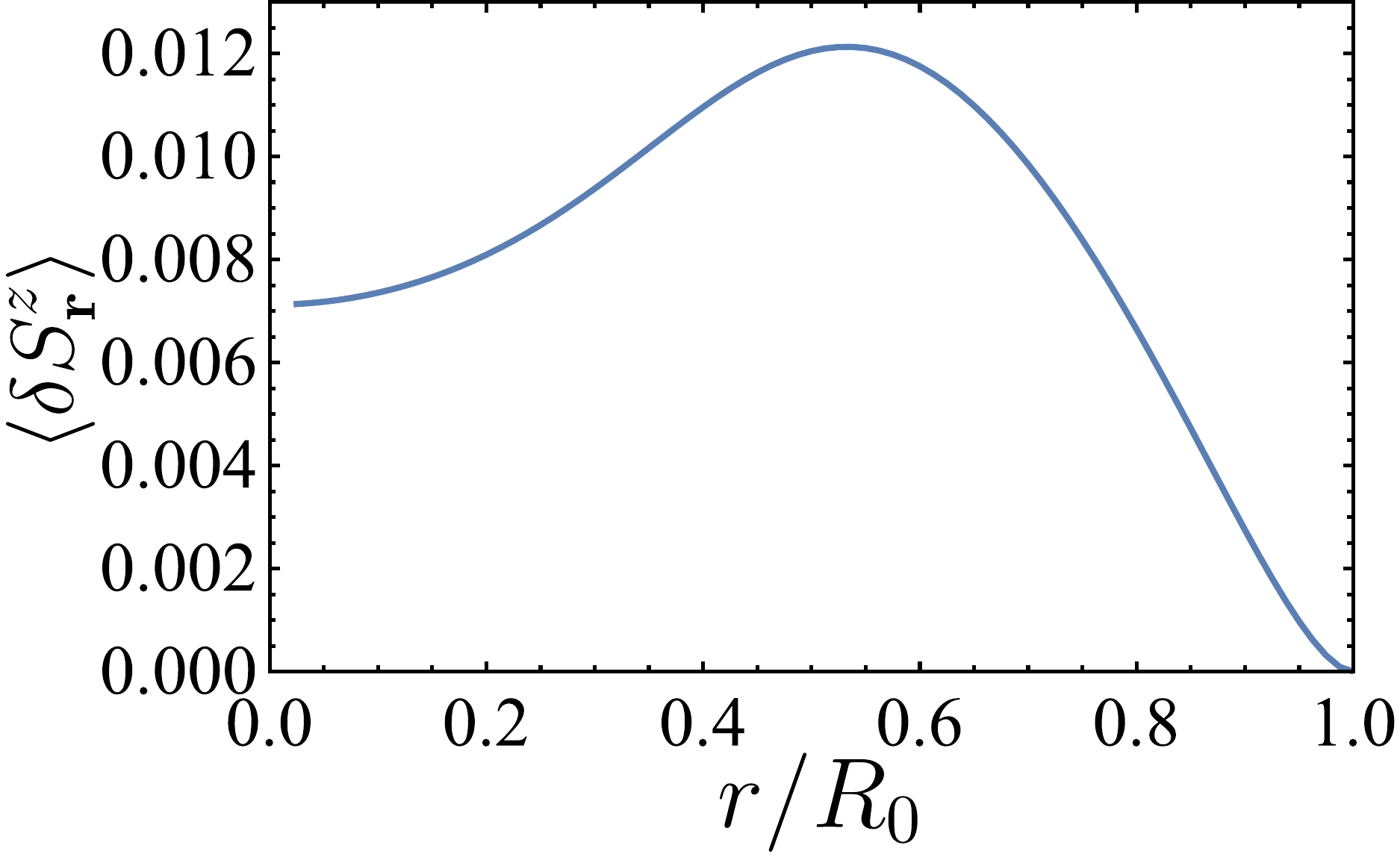}
\caption{The reduction of equilibrium spin value due to zero-point motion shown as a function of distance to the center of the disc. }
\label{fig:spin_reduction}
\end{figure}

Using the above formulas \eqref{uvfunction} we represent the spin reduction in the form 
\begin{equation}
\langle a^{\dagger} _{\mathbf{r} } a _{\mathbf{r} }\rangle  = 
\sum_{mk} \left | {\tilde \phi}_{k}^{(m)}( r )  \right|^{2}
\end{equation}
The obtained quantity does not depend on $s$ and is plotted in Fig.\ \ref{fig:spin_reduction}. It is seen that even for $s\sim1$ the contribution of zero-point motion is negligible. 

\section{Band structure of magnons  on skyrmion lattice
\label{sec:bands}}

Until now we discussed the spectrum of magnons confined on the disc with the skyrmion configuration of the classical state. The demand for the spin orientation at the edge be parallel to the field, $\beta(R) =0$, led to the property $\psi_{m,n}(R)=0$. The latter condition may be modeled by the quantum box potential $U_{edge}(r =R)=\infty$. When discussing   the magnon spectrum for the skyrmion lattice we first pave the whole plane by the hexagons, which are separated from each other by the infinite edge potentials. The height of these potentials is then gradually reduced to zero, which results i) in the continuum spectrum for delocalized magnon states with $E>b$ and ii) in the tight-binding form of the spectrum for states with energies $E<b$. 

Strictly speaking, the estimate for the boundary for the continuum spectrum, $E_{c}=b$, is obtained from \eqref{eq:HamQuantBose} by putting there $\beta(R)=0$ and neglecting derivatives,  $\beta'(R)$, in which case the Hamiltonian is diagonal and corresponds to free motion with the dispersion $E_{k} = C k^{2} +B$. It turns out that for the disc of optimal radius the derivatives, $\beta'(R)$, are sizable and the picture becomes complicated as discussed below. For our estimates it is enough to take  $E_{c} \simeq b$.   

\begin{figure}[t]
\includegraphics[width=0.95\columnwidth]{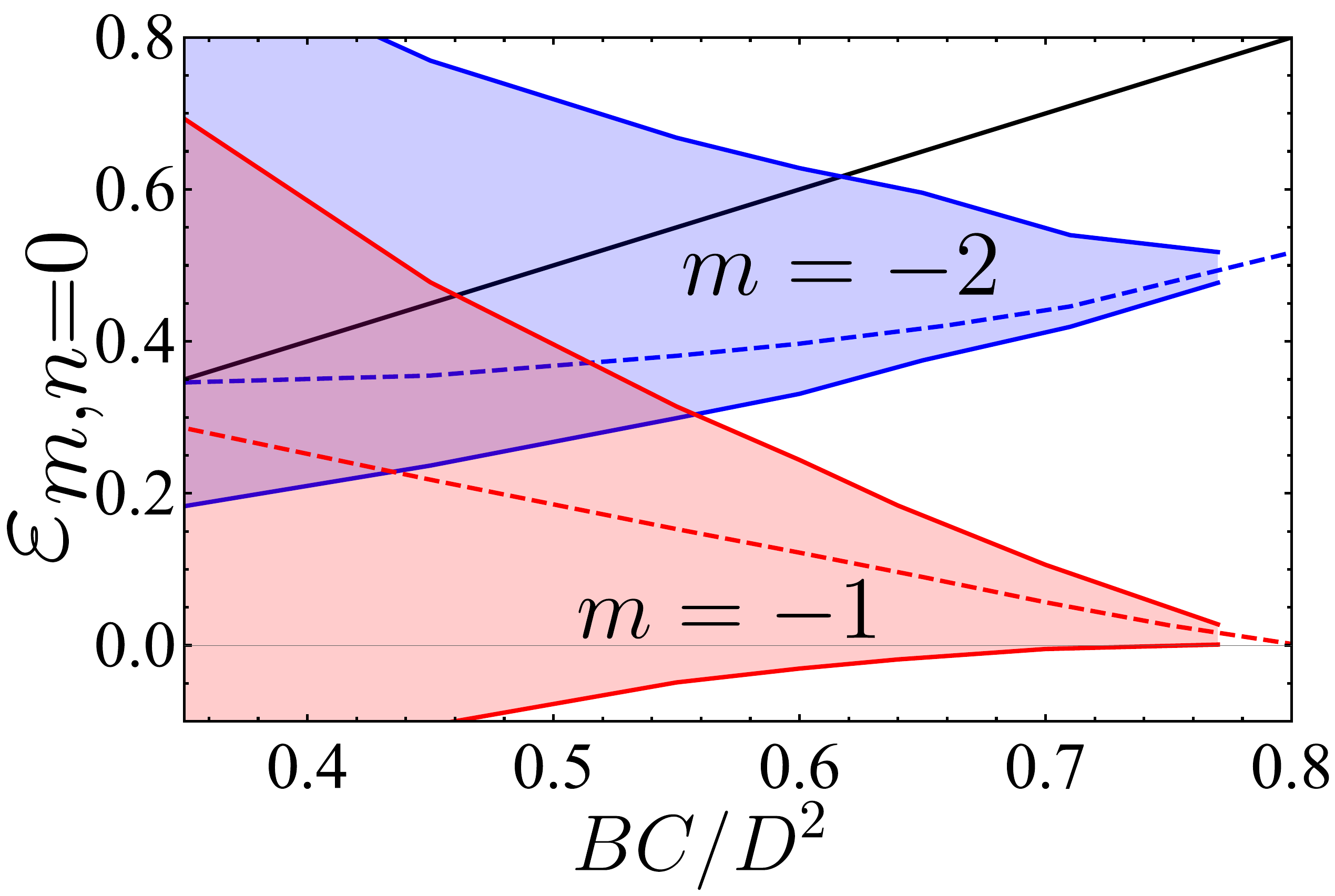}
\caption{\label{pic:2lowest} 
Dependence of the energy of two lowest modes $\epsilon _{m=-1,0}$ and $\epsilon_{m=-2,0}$ on the magnetic field,  $b$, calculated  for the optimal radius of the disc $R_{0}$. The energies for the top and the bottom of the bands is showed too. Notice the almost vanishing value of $\epsilon_{m=-1,0}$ at  $b\simeq 0.8$. The estimated boundary of spin-wave continuum is shown by straight solid line.}
\label{fig:LatticeDispersion}
\end{figure}

In the Fig. \ref{pic:plotModeEnerg} we see that for optimal radius of the disc only two magnon states with $m=-1$ and $m=-2$ lie below the magnon continuum $E < E_{c} \simeq 0.6$. 
We show the evolution of these energy modes with field in the Fig.\ \ref{pic:2lowest}.
The wave function of these states is mostly present at the center of the disc and is relatively small at its edge. When lowering the confining barrier between discs (hexagons) these states should remain almost localized we expect that the dispersion obtains the tight-binding form : 

\begin{equation}
\begin{aligned}
    \varepsilon_{m} (\mathbf{k}) & = \varepsilon_{m,n=0}^0 
 + 6 t_{m} \gamma(\mathbf{k}) \,, \\ 
 \gamma(\mathbf{k}) & = \tfrac 13
 \big( \cos  k_{x}\ell 
 + 2\cos  \tfrac12 {k_{x}\ell}  \, \cos  \tfrac{\sqrt{3}}2 k_{y} \ell   \big)   
\end{aligned}
\label{eq:dispersionOfEnergy}
\end{equation}
with $\ell \simeq 1.94 R_{0}$ is the distance between neighboring skyrmions, obtained from equality of the area of  disc and hexagon, $\pi R_{0}^{2} = \sqrt{3}\,\ell^{2}/2$. The highest and the lowest band energies are $ \varepsilon_{m,n=0}^0 + 6t_{m}$ and  $ \varepsilon_{m,n=0}^0 -3 t_{m}$  which is  achieved at $\mathbf{k}=0$ and $\mathbf{k} = (4\pi/3\ell,0)$, respectively.  Our aim now is to determine the on-site energy $\varepsilon_{m,n=0}^0$ and the hopping amplitude  $t_{m}$. We estimate these quantities by evaluating the wave-functions at the edge. 


Consider first a usual example of 1D symmetric two-well potential with $H = -\nabla^{2} + U(x)$, whose low-lying states $\epsilon _{i}$ lie below the barrier potential $U(0)$. In the semiclassical regime the wave-function of the individual well decreases exponentially at the barrier $\psi(x\simeq 0) \sim \exp (-\kappa x) $, with $\kappa^{2} \simeq U(0)-\epsilon _{i}$. The eigenfunctions are obtained as symmetrized combinations $\psi_{s} \sim \cosh(\kappa x)$, $\psi_{a} \sim \sinh(\kappa x)$. The   solution $\psi_{a} $ has lower energy and the energy difference is   $\delta E = E_{s}-E_{a} \simeq 2 \psi_{s}(0) \psi_{a}'(0) =2 (\psi_{a}'(0))^{2}/\kappa$.  \cite{Landau1981Quantum}  

In comparison with this usual case we have three modifications. First we have to pass from the 2D Hamiltonian, $r^{-1}\partial _{r} r \partial _{r}$,  to 1D case, $\partial_{r}^{2}$ by making substitution $\psi(r) = r^{-1/2}\tilde \psi (r)$. Second, more importantly, we have a two-component spinor form of wave functions, rather than scalar one. As a result, we have two eigenfunctions $\psi_{1,2}$ at the edge, characterized by two values, $\kappa_{1,2}$.  Third modification addresses the parity of the wave-function in the individual well. For odd functions the lower energy combination $\psi_{a}$ is given \emph{by the sum} of individual functions. In our case it happens for odd $m$, and is accompanied by a change of sign of $t_{m}$ in \eqref{eq:dispersionOfEnergy}.   

We proceed by linearizing the equations \eqref{eq:ShrodEq} at $r=R$  
\[
( -\partial^{2}_{r} + \hat M - \tau_3 \varepsilon _{m,0} ){ \tilde \psi }_{m,0} =0 
\]
with $m=-1,-2$ and constant $\hat M = {\hat {\cal H}}^{(m)}_{r=R}$. The above equation has  general solution ${ \tilde \psi }_{m,0} (r) = \sum_{i=1,2} c_{i} \tilde \psi_{i}^{0} \sinh \kappa_{i} (r-R) $, having the property ${ \tilde \psi }_{m,0}(R)=0 $. Knowing the eigenfunctions ${  \psi }_{m,0}(r)$ of the full  equation \eqref{eq:ShrodEq} we can extract the weights $c_{i}$ in ${ \tilde \psi }_{m,0} (r)$  and verify that this function is a good approximation for the exact numerical  $r^{1/2} {\psi }_{m,0}(r)$ at $r\simeq R$.  Finally the energy splitting is found as  
$ \delta E = 2 \sum_{i=1,2} c_{i}^{2}   \kappa_{i} \langle \tilde \psi_{i}^{0} | \tilde \psi_{i}^{0} \rangle >0$. 

Denote the lowest energy level as $E_{1} \equiv \varepsilon _{m=-1,n=0} $.   In our approximation by the two-well potential the symmetrical solution has the energy  $E_{s} = E_{1} + \delta E_{1}$ and hence  the on-site energy is $E_{1}+\tfrac12 \delta E_{1} $ whereas the hopping is $t_{1}= -\delta E_{1} /2$, with the change of sign discussed above. It follows that the bottom of the lowest band on the hexagonal superlattice has the energy $E_{bottom} = E_{1} - \tfrac52 \delta E_{1}$, and the top of the band corresponds to $E_{top} = E_{1} + 2 \delta E_{1}$. Similar consideration holds for $m=-2$, but now $t_{2}= + \delta E_{2} /2$. 

 If our approximate calculation were exact, we would have $E_{bottom} =0$ for $m=-1$.  This property is expected, since the finite energy  $\varepsilon _{m=-1,n=0}$ corresponds to a broken translational symmetry on the disc, and consideration of the skyrmion lattice restores the translational symmetry and should result in the Goldstone mode. 
 
We show the energies for the top and the bottom of the bands with $m=-1$, $m=-2$ in the Fig. \ref{fig:LatticeDispersion}.
Both bands become narrow near critical field $b\simeq 0.8$, because the distance between the skyrmions  increases and the overlap of the individual wave-functions decreases in this case. In this case the semiclassical treatment is justified and we see  that  $E_{bottom}$  touches zero for $m=-1$, as expected. 
However, $E_{bottom}$ becomes negative for smaller values of $b$, where the semiclassical regime fails;  it provides the estimate for the accuracy of our calculation. 

 At lower fields, $b\alt 0.5$   the top of the lower bands increases to magnon continuum at $E\sim b$.  In this case our initial picture of localized levels in individual wells and their eventual hybridization is hardly justified. This argument can be applied already to Fig.\ \ref{pic:2lowest} with the lower bound on the field $b\simeq 0.35$, and our calculation of the bandwidths changes this bound to higher values,  $b\sim 0.5 \div 0.6$. It means that the skyrmion crystal can be considered as such in the range of the fields, $b \in (0.6,0.8)$. Similar estimate is found in \cite{Schutte2014}, however based on the other arguments. 
 
\subsection{3D case 
\label{sec:3d}}

 The generalization of our analysis to the three-dimensional case is straightforward. We assume that the classical configuration of spins is given by 2D picture extended to third dimension. Instead of points on the plane, representing the centers of skyrmions, we now have the vortex lines in space. The classical treatment of Sec.\ \ref{sec:lattice} is unchanged. The consideration of Sec. \ref{sec:spectrum} is modified as follows.  The  Schr\"odinger equation \eqref{eq:HamQuantBose} acquires the term $-\nabla_{z}^{2}$   and is   solved by separation of variables. The basis functions in  \eqref{eq:expansionBose} are $e^{ik_{z} z} e^{im\varphi } f_{n}^{(m)}( r )$ with the resulting change in the spectrum  
 \begin{equation}
 \varepsilon^0_{m,n} \to \varepsilon^0_{m,n} + C k_{z}^{2} \,,
 \label{spectrum3D}
 \end{equation} see Eq.\eqref{eq:numMatrixEq}. The conclusions of the rest of this section are unchanged, including the discussion around Eq.\ \eqref{spiral_energy}.  The analysis on the lattice, Sec. \ref{sec:bands}, now includes the  modification of the spectrum \eqref{eq:dispersionOfEnergy} according to \eqref{spectrum3D}, which does not change the rest of this section. 
 
What is important, is the possibility to discuss  in 3D case the case of non-zero temperature without losing the long-range order. In purely 2D case the calculation of the spin reduction gives divergent quantity. We may estimate  
$$\langle a^{\dagger} _{\mathbf{r} } a _{\mathbf{r} }\rangle - 
\langle a^{\dagger} _{\mathbf{r} } a _{\mathbf{r} }\rangle _{T=0} \sim  \sum _{m}  
   \int d\mathbf{k}  N(\varepsilon_{m} (\mathbf{k}))$$ 
 here  $N(\omega) = (\exp(\omega/T)-1)^{-1}$, and we ignored the amplitude of wave-functions for simplicity. The singular contribution to averaged fluctuations comes from the Goldstone mode $m=-1$. 
At small $|\mathbf{k} |$ we have  $\varepsilon_{-1} (\mathbf{k}) \propto  k^{2} $ in  \eqref{eq:dispersionOfEnergy}, which leads to logarithmic divergence at $T\neq 0$ and a loss of magnetic order in skyrmion lattice, in accordance with Mermin-Wagner theorem.  The third spatial dimension   improves the situation, as we may read from the Fig.\  \ref{fig:LatticeDispersion}   for intermediate fields 
 \begin{equation}
\begin{aligned}
 \varepsilon_{-1} (\mathbf{k})  & = C k_{z}^{2} +  \varepsilon_{-1,0}^0 (1-  \gamma(\mathbf{k}))
\simeq     C    k_{z}^{2}  + C_{plane} k^{2}   \,,
 &    \\
C_{plane} &  \sim    (0.8 - b) C  \,. \\
\end{aligned}
\label{spectrum3D.low}
\end{equation}
Notice here that $ \varepsilon_{-1,0}^0 \sim D^{2}/C$ and  $1-  \gamma(\mathbf{k}) \sim (k\ell)^{2}$ with $\ell \sim C/D$, when we restore dimensional units. 
The integration over in-plane $\mathbf{k}$ is done within the hexagonal Brillouin zone, $|\mathbf{k}|\alt \ell^{-1}$, related to skyrmion lattice with a period $\sim C/D$.  Simple estimate gives for this singular contribution 
\[ \delta \langle a^{\dagger} _{\mathbf{r} } a _{\mathbf{r} }\rangle \sim \frac{T D}{C^{2}\sqrt{0.8-b}} \, \]
which is small except for $b\simeq 0.8$. 
The contribution from the bands with $m\neq-1$ can be estimated by approximate equating it to the case of uniform ferromagnet  
 \[  \langle a^{\dagger} _{\mathbf{r} } a _{\mathbf{r} }\rangle \sim  \int d^{3}\mathbf{k}  \frac{T}{C k^{2} + B} \sim \frac TC\]
Summarizing here, we showed that the long-range order is stabilized in 3D and is unaffected by the existence of translational Goldstone mode of the skyrmion crystal everywhere except the close vicinity of the critical field.  
  
 \section{Conclusions
 \label{sec:conclusions}}

We calculated the spectrum of magnons in the skyrmion crystal by first solving the eigenvalue problem on the disc of finite radius and then considering the hybridization of the discrete energy levels in the hexagonal skyrmion superlattice. Following the earlier works, we determine the superlattice spacing by minimization of classical energy density.   The finite  superlattice spacing leads to stability of the system at the level of one cell, and only two energy levels lie below the magnon continuum and may be considered as localized.  The subsequent hybridization of these levels leads to the bands of finite width in the hexagonal tight-binding scheme. We show that the translational Goldstone mode is approximately restored in our numerical approach. The almost-localized character of the lowest energy levels is lost at smaller values of the field due to significant overlap of these lower bands with the higher-energy continuum. At higher fields $B\simeq 0.8 D^{2}/C$ the superlattice period diverges and the quantum fluctuations destroy the long-range order. This may be interpreted as a disappearance of the skyrmion crystal  at the fields $B$ outside the interval $(0.5 \div 0.8)  D^{2}/C$. The classical energy density of the skyrmion lattice lies slightly above the energy of the single conical spiral in the whole range of magnetic fields.  The calculated quantum correction  of order $1/s$ to the classical energy can favor the skyrmion crystal at higher temperatures, when the average value of spin is reduced.  Our findings may indicate the mechanism for creation of skyrmion crystal in the so-called A-phase in B20 compounds observed at high temperatures in the restricted range of magnetic fields. \cite{Muhlbauer2009}

 \acknowledgements

We thank  M. Garst,  S.V. Maleyev, K.L. Metlov,  A.O. Sorokin  for useful discussions. 

\appendix
\section{\label{app:discrete} From lattice model to continuum model  }

In this section we derive the continuum version of our model, suitable for possible subsequent analysis of the interaction between magnons.   We use below Latin indices for spin space coordinates ($a,b,c,d = 1,2,3$), and Greeks indices for physical space ($\alpha, \beta, \gamma = 1,2$). We write the 
Hamiltonian   (\ref{eq:ham}) in the form
\begin{equation}
\sum\limits_{\mathbf{r},\mathbf{n}} \left\{ S_{\mathbf{r}}^a\left( J\left( \mathbf{n}\right)\delta _{ab} 
+ D\varepsilon _{abc}\delta _{\alpha c} n^\alpha \right)S_{\mathbf{r} - \mathbf{n}}^b \right\}  - sB\sum\limits_{\mathbf{r}} S_{\mathbf{r}}^a\delta _{a3}
\label{eq:mainHam}
\end{equation}
with  totally antisymmetrical tensor $\varepsilon_{abc}$ and magnetic field $B$ supplied by factor $s$ for further convenience.  DM interaction has mixed spin and space indices stemming from its spin-orbital nature.

Let the ground state be characterized by non-collinear spin configuration. We express Eq. (\ref{eq:mainHam}) in such local basis, where the average local spin, $\tilde {\mathbf{S}}_{\mathbf{r}}$,  is directed along the $\hat z$-axis, with $\mathbf{S}_{\mathbf{r}}  = \hat U\left( \mathbf{r} \right ) \tilde {\mathbf{S}}_{\mathbf{r}}$.
The position-dependent matrix, $\hat U(\mathbf{r}) = e^{-\alpha \sigma_3}e^{-\beta \sigma_2}e^{-\gamma \sigma_3}$ is defined with generators of $SO(3)$ group $\sigma_2$, $\sigma_3$ and Euler angels $\alpha$, $\beta$, $\gamma$.

The Euler angle $\gamma$ is not determined by variational equation (\ref{eq:eulerLagr}) and we may chose  $\gamma = \alpha$ for the continuity of LSWT equation \eqref{eq:HamQuantBose} at $r=0$, see Ref.\ \cite{Aristov2015}.    In terms of bosons the choice of $\gamma$ is encoded in simple $U(1)$ transformation, $a \to e^{ - i\gamma }a$, and we adopt $\gamma = 0$ in this paper, in order to make a better comparison with other authors \cite{Schutte2014}. 

We assume the long wavelength limit, ${\mathbf{qn}} \ll 1$, which particularly corresponds to smooth variation of $ \hat U\left( \mathbf{r} \right )$ on the scale of interatomic distances. In this case we can write 
\[ S_{\mathbf{r} + \mathbf{n}}^b  \simeq (1+ n^{\alpha} \nabla^{\alpha} + n^{\alpha} n^{\beta} \nabla^{\alpha} \nabla^{\beta} ) (\hat U\left( \mathbf{r} \right ) \tilde {\mathbf{S}}_{\mathbf{r}} ) \]
It is convenient to define quantities
\begin{equation}
\begin{aligned}
  \chi _{1,\alpha }^{ab} & =U^{ca}\nabla ^\alpha U^{cb},    \\
  \chi _{2,\alpha \beta }^{ab} & = U^{ca}\left( \nabla ^\alpha \nabla ^\beta U^{cb} \right)   \\ 
\end{aligned}
\label{eq:chi}
\end{equation}
with the explicit expressions for $U(\mathbf{r})$ and $\chi_1(\mathbf{r})$, $\chi_2(\mathbf{r})$ given in \cite{Aristov2015}. 
We  perform  the summation over $\mathbf{n}$ according to the rules 
\begin{equation*}
\begin{aligned}
  \sum\limits_{\mathbf{n}} n^\alpha J(\mathbf{n}) & = 0  \,, \quad 
  \sum\limits_{\mathbf{n}} n^\alpha n^\beta J(\mathbf{n})  
  =  - \left. \frac{d^2J(\mathbf{q})} {dq^\alpha dq^\beta}\right|_{q = 0}  \,,  \\ 
   \sum\limits_{\mathbf{n}} \varepsilon _{abc}\delta _{\alpha c}n^\alpha   & = 0    \,, \quad
  \sum\limits_{\mathbf{n}} \varepsilon _{abc}\delta _{\alpha c}n^\alpha n^\beta  
   = \varepsilon _{abc}\delta _{\beta c}    \,,
\end{aligned}
\label{eq:integrExch}
\end{equation*}
with $
J(\mathbf{q}) = \sum\limits_{\mathbf{n}} e^{i\mathbf{qn}}J(\mathbf{n})  \simeq J(0) + \tfrac{1}{2}Cq^2 \,.
$
 As a result we come to the Hamiltonian 
\begin{equation}
\begin{aligned}
H  &= \int d\mathbf{r}\, ( H_{ex} + H_{DM} + H_B ) \,, \\ 
H_{ex} &=  - \frac{1}{2}C \, 
 {\tilde S}_{\mathbf{r}}^a \left( \chi _{2,\beta \beta }^{ab} + 2\chi _{1,\beta }^{ab} \nabla ^\beta 
 + \delta _{ab}\Delta  \right)\tilde S_{\mathbf{r}}^b \,, 
\\
H_{DM} &=  - D \, \tilde S_{\mathbf{r}}^a\varepsilon _{adc}\delta _{e\alpha }U^{ec}
\left( \chi _{1,\alpha }^{db} + \delta _{db}\nabla ^\alpha  \right)\tilde S_{\mathbf{r}}^b  \,,
\\
H_ B  &=  - sB \, U^{3a}\tilde S_{\mathbf{r}}^a \,.
\end{aligned} 
\label{eq:hamlitNewBasis}
\end{equation}
The classical part of the energy, Eq.\ \eqref{eq:en_class}, is obtained by putting $\tilde{ \mathbf{S}}_{\mathbf{r}} = (0,0,s)$ in Eq.\ \eqref{eq:hamlitNewBasis} and using the convention \eqref{eq:parametr}. 
Explicit formulas are given in Eq.\ \eqref{eq:ClassGlobalForm}. 
 
The terms linear in bosons have the factor $s^{3/2}$ and vanish due to the extremum condition on $\beta$ (\ref{eq:eulerLagr}) after integration by parts. Quadratic terms in bosons lead to Eq.\ \eqref{eq:HamQuantBose}.

\section{\label{app:num} Numerical methods}

The eigenfunctions to (\ref{eq:ShrodEq}) are expanded in the basis of the orthonormal functions 
\begin{equation}
f_{k}^{(m)}\left( \mathbf{r} \right) = 
\frac{\sqrt 2 }{R_0 \, | J'_{m+1}( z_k^{(m+1)}  ) |}J_{m+1}\left( z_k^{(m+1)}\frac{r}{R_0} \right)
\label{eq:basisMain}
\end{equation}
with $k=0,1,2\ldots$,  $J_m(x)$ the Bessel function and $z_k^{(m)}$ its $(k+1)$-th positive zero. The set $\eta_{mk} = e^{i(m+1)\varphi } f_{k}^{(m)}$ satisfies the Laplace equation on a disc of radius $R_0$ and with boundary conditions 
\[ 
 - \nabla^{2} \eta_{mk} = \lambda _k^{(m + 1)} \eta_{mk}, 
 \quad \eta_{mk} (R_0) = 0, \quad \eta_{mk} (0) < \infty
\]
with eigenvalues $\lambda _k^{(m + 1)} = (  {z_k^{(m+1)}}/ {R_0} )^{2}$.

%

In the above basis the equation (\ref{eq:ShrodEq}) acquires the matrix form
\begin{equation}
\sum_{k\geq0}^{\infty}
 \begin{pmatrix}
F_{nk} (m)&G_{nk}\\
 - G_{kn} & - F_{nk} (-m)
 \end{pmatrix}
\begin{pmatrix}  u_k^{(m)}\\    v_k^{(-m)}  \end{pmatrix}  
 = \varepsilon^0_{m,n}  \begin{pmatrix}  u_n^{(m)}\\    v_n^{(-m)}  \end{pmatrix} 
\end{equation}
with vectors  $\varepsilon^0_{m,n}$ and $u_n^{(m)}$, $v_n^{(-m)}$  of, generally speaking,  semi-infinite length. 
Practically, it suffices to take the basis of 40 lowest $z_{k}^{(m)}$'s to achieve a good accuracy for the low-energy spectrum.  The  matrix elements are given by 
\begin{equation}
\begin{aligned}
F_{nk} \left( m \right) &=  \lambda _n^{(m + 1)}{\delta _{nk}} 
\\ & + \int rdr {\left[ V_0 +  m V_1 \right] f^{(m)}_{n} \left( r \right) f^{(m)}_{k} \left( r \right)}  \\ 
G_{nk}  &= \int rdr \left[ C\left( \frac{{{{\sin }^2}\beta }}{{2{r^2}}} - \frac{1}{2}{\left( \frac{d\beta}{dr} \right)}^2 \right) \right.    \\
& \left. + \quad D\left(\frac{\sin 2\beta}{2r} - \frac{d\beta}{{dr}}\right) \right]  f^{(m)}_{n} ( r  ) f^{(-m)}_{k} ( r ) 
\end{aligned}
\end{equation}
with 
\begin{equation}
\begin{aligned}
V_0 & =  - \frac C2 \left( \frac{3  \sin^{2}\beta}{ r^2}   +\left( \frac{d\beta }{dr} \right)^2 \right)   \\
& - D\left( \frac{3\sin2 \beta }{2r} + \frac{d\beta }{dr} \right) + b\cos \beta  \,, \\
V_1 & =  \frac{2}{r}\left( D\sin \beta  - \frac{C}{r} ( \cos \beta  + 1  ) \right) \,.
 \end{aligned} 
\end{equation}
The spectrum of magnons  is numerically  obtained by diagonalizing  the  matrix (\ref{eq:numMatrixEq}).
We calculate the spectrum for $m>0$ only, due to the property $ \tau _1 \tau_{3}{\widehat H^{(m)}} \tau _1 =  - {\widehat H^{( - m)}}$, with Pauli matrices $\tau_{i}$.     The positive eigenvalues then give $\varepsilon^0_{m,n}$ for $m>0$ and negative eigenvalues correspond to  $-\varepsilon^0_{m,n}$ for $m<0$.

\section{\label{app:spiral}  Energy of single spiral}

In this section we evaluate the classical energy of the single conical spiral, whose form is given by the expression \eqref{eq:parametr} with 
\begin{equation}
\alpha = \mathbf{qr} \,,\quad \beta = const \,.
\end{equation}
Here  $\mathbf{qr} =  q_{x}x +  q_{y}y$, with vector $\mathbf{q}$ determined shortly.  
The expression \eqref{eq:parametr}  is written in orthogonal frame $\hat e_{1}$,  $\hat e_{2}$,  $\hat e_{3}$, with angle $\Theta$ between $\hat e_{3}$ and normal to the plane, $\hat e_{z}$,  i.e.\  $\hat e_{3} \dot \hat e_{z} = \cos \Theta$.  Let the field $\mathbf{B}$ be directed at an angle to the plane, $\mathbf{B} \hat e_{z} = B \cos \Theta_{1} $, and to the center of the cone, $\mathbf{B} \hat e_{3} = B \cos \chi $. Simple calculation with the use of  \eqref{eq:en_class} gives for the energy density 
\begin{equation}
\rho = \frac{C}2   q^{2} \sin^{2}\beta - \mathbf{qd}   \sin^{2}\beta - B \cos\chi \cos \beta \,,
\end{equation}
where $\mathbf{d} = D( \sin \Theta, - \cos\Theta, 0 )$ in the laboratory frame 
and we chose $\hat e_{3} \dot \hat e_{y} = 0$ for definiteness.  Clearly, the minimum of $\rho$ happens at $\mathbf {q}=   \mathbf{d}/C$ and at   $\chi =0$. The latter condition shows that the cone axis is directed along the field, so that $\Theta_{1}= \Theta$ ;  we obtain
\begin{equation}
\rho_{sp} =  - \frac{D^{2}}{2C}  \sin^{2}\beta - B  \cos \beta \,.
\end{equation}
Variation over $\beta$ gives $\cos\beta = BC/D^{2} = b $ for $b<1$ and $\cos\beta=1$ otherwise.
Subtracting here the energy of uniform ferromagnet $\rho_{FM} = -B$, we obtain the  expression \eqref{spiral_energy} for the energy gain of the single spiral.
\cite{Maleyev2006}


 \end{document}